\documentclass[11pt,a4paper]{article}

\usepackage[russian,english]{babel}
\usepackage{amscd}
\usepackage{amsmath}
\usepackage{amsfonts}
\usepackage{amssymb}
\usepackage{graphics}
\usepackage{epsfig}
\usepackage{array}  
\usepackage{longtable} 
\usepackage[ruled,vlined]{algorithm2e}

\usepackage{subcaption}
\newcommand{\CS}{C\#} 
\newcommand{\pluseq}{\mathrel{+}=}

\newcommand\abs[1]{\left|#1\right|}
\DeclareMathOperator{\sign}{sign}
\DeclareMathOperator{\modulo}{mod}
\DeclareMathOperator{\mathif}{if}
\DeclareMathOperator{\mathscale}{scale}
\DeclareMathOperator{\mathrotation}{rotation}
\DeclareMathOperator{\mathtranslation}{translation}

\DeclareMathOperator{\mathmax}{max}
\DeclareMathOperator{\mathdot}{dot}
\DeclareMathOperator{\mathlerp}{lerp}
\DeclareMathOperator{\mathsaturate}{saturate}

\SetKwProg{Fn}{Function}{}{end}
\SetKwInOut{Input}{Input}
\SetKwInOut{Output}{Output}

\title{SN-Engine, a Scale-free Geometric Modelling Environment}
  
\author{T.A.\,Zhukov}

\date{}

\begin{document}

\newcommand{\url}[1]{#1}

\maketitle

%

\begin{abstract}

We present a new scale-free geometric modelling environment designed by the author of the paper.
It allows one to consistently treat geometric objects of arbitrary size and offers extensive analytic 
and computational support for visualization of both real and artificial sceneries.

\medskip

\end{abstract}



\section{INTRODUCTION} 

Geometric modelling of real-world and imaginary objects 
is an important task that is ubiquitous in modern computer science. 
Today, geometric modelling environments (GME) are widely used 
in cartography~\cite{RCART}, architecture~\cite{RARCH}, geology~\cite{RGEO}, hydrology~\cite{RHYDRO}, and astronomy~\cite{RASTRO}.
Apart from being used for modelling, importing and storing real-world data, 
GMEs are crucially important in computer games industry~\cite{PROCUENV}.

Creating a detailed model of a real-world or imaginary environment 
is a task of great computational complexity. Modelling various 
physical phenomena like gravity, atmospheric light scattering,  
terrain weathering, and erosion processes requires full use of modern computer algebra algorithms 
as well as their efficient implementation. 
Finding exact and numeric solutions to differential,
difference, and algebraic equations by means of computer 
algebra methods (see~\cite{BASES} and the references therein) is at the core of the crucial algorithms 
of modern geometric modelling environments.

There exist a wide variety of geometric modelling environments,
both universal and highly specialized. They include three-dimensional planetariums 
like SpaceEngine or Celestia, virtual Earth engine Outerra, procedural scenery generator Terragen,
3D computer graphics editors Autodesk 3ds Max, Blender, Houdini, and 3D game engines like
Unreal Engine 4, CryEngine 3, and Unity.

The present paper describes a new geometric modelling environment, 
SN-Engine, designed and implemented by the author of the paper. The main advantage of this GME is its capacity to consistently 
and in a computationally efficient way treat 
geometric objects of arbitrary size, from extragalactic  
and down to the microscopic scale.

This geometric modelling environment is a freeware implemented in \CS\, programming language. Scripts, sample video 
files and high resolution images created with SN-Engine are publicly available at snengine.tumblr.com.

%


\section{STATE-OF-THE-ART IN GEOMETRIC MODELLING: CAPABILITIES AND LIMITATIONS} 
The geometric modelling environment presented in the paper shares 
many properties and functions 
with other GMEs. These include procedural generation, instruments and
algorithms of 3D rendering, internal elements of system architecture,
file and in-memory data formats, etc. In the next tables we
compare SN-Engine with other existing systems.

\begin{table}[ht!] \scriptsize \centering
\caption[Space modelling systems]{Space modelling systems} 
\label{tab:t1}
\begin{tabular}[]{@{}|p{2.2cm}|p{2.5cm}|p{3.2cm}|p{3.8cm}|@{}}
\hline
\textbf{Title} & \textbf{Type} & \textbf{Procedurally generated objects} & \textbf{Planet surfaces}\\ \hline 
SpaceEngine    & astronomy program, game engine& galaxies, star systems, planets          & detailed surface, no ground objects \\ \hline 
Elite Dangerous& game                          & galaxy, star systems, planets            & detailed surface, no atmospheric planets \\ \hline  
Star Citizen   & game, publicly available alpha version & planet details, object layouts  & detailed surface, ground objects, no true-scale planets\\ \hline  
No Man's Sky   & game                          & planets, objects, species                & detailed voxel terrain, ground objects, no true-scale planets\\ \hline  
Celestia       & astronomy program             & none                                     & textured model \\ \hline 
Outerra        & 3D planetary engine           & planet surface details, objects          & detailed surface, biomes, Earth data, ground objects \\ \hline 
SN-Engine        & world platform/game engine    & galaxies, star systems, planets, objects & detailed surface, biomes, Earth data, ground objects \\ \hline 
\end{tabular}
\end{table} 
\begin{table}[ht!] \scriptsize \centering
\caption[3D computer graphics editors]{3D computer graphics editors} 
\label{tab:t2}
\begin{tabular}[]{@{}|p{1.6cm}|p{3cm}|p{4.2cm}|p{2.9cm}|@{}}
\hline
\textbf{Title}& \textbf{License } & \textbf{Use} & \textbf{Type of procedural generation} \\ \hline 
Blender          & free & 3D computer graphics, mixed                              & scripted             \\ \hline 
Autodesk 3ds Max & commercial & 3D computer graphics, mixed                              & scripted             \\ \hline   
Autodesk AutoCAD & commercial & computer-aided design and drafting                             & partial, scripted    \\ \hline   
Houdini          & commercial & 3D animation, mixed                              & integrated, scripted \\ \hline   
CityEngine       & commercial & urban environments generation      & integrated           \\ \hline   
Terragen         & commercial, free & procedural landscape generation and visualization & integrated           \\ \hline   
SN-Engine          & no public release yet & procedural world creation          & integrated, scripted \\ \hline   
\end{tabular}
\end{table} 
\begin{table}[ht!] \scriptsize \centering
\caption[3D game engines]{3D game engines} 
\label{tab:t3}
\begin{tabular}[]{@{}|p{1.6cm}|p{2.5cm}|p{2.5cm}|p{2.5cm}|p{2.5cm}|@{}}
\hline 
\textbf{Title} & \textbf{Game creation methods} &\textbf{Game creation\newline license} & \textbf{Source\newline availability} & \textbf{Type of procedural generation }\\ \hline 
Unreal Engine 4 & editor, visual programming, C++ programming  & free / royalty for commercial use  & open                                 & plugins, scripted\\ \hline   
Unity           & editor, \CS\,programming, prefab construction & free / subscription model          & closed / open on enterprise license & plugins, scripted\\ \hline   
CryEngine 3     & editor, C++ programming                      & free / royalty for commercial use  & open                                 & plugins, scripted\\ \hline   
SN-Engine         & editor, Lua programming, \CS\,programming     & free                               & open scripts, closed source code    & intergated, scripted\\ \hline   
\end{tabular}
\end{table}

Despite the extensive capabilities of modern geometric modelling systems 
all of them have 
limitations in terms of world structure, engine
modification, user-world interaction, and licensing.
The presented modelling system SN-Engine is a freeware which, unlike
any other software listed above, is endowed with tools for procedural 
generation of objects of arbitrary scale.


The presented geometric modelling system unifies procedural approach to content generation 
and game engine technologies to create a fully flexible 
multi-client, arbitrary scale world creation and experience 
platform. The system can be used for solving various tasks like 
demonstration, design, recreation, and education. Its functions 
include generation of new content, transformation of existing content, 
and visualization. By organizational structure the presented geometric modelling
system can be used in autonomous, local-networked or global-networked mode.


The engine provides flexibility in terms 
of world and entity modification, creation and generation.
The list of main features is as follows:
\begin{itemize}  
	\item Support of arbitrary scale worlds, from super galactic to microscopic level;
	\item Client-server world and logic synchronization;
	\item Scriptable game logic;
	\item Procedural generation~\cite{PROCUENV,PROCCAVE,TERGEN,PROCINT} of any world content ranging 
	from textures and models, to full world generation;
	\item External data import: e.g. height maps~\cite{DEM,TERRD}, OpenStreetMap~\cite{OSM}
	data, textures, models, etc.;
	\item Integrated computational physics module;
	\item World state system for saving world data including 
	any runtime changes;
	\item Script system controlling every world	related 
	function of the engine; 
	\item Fully extendable and modifiable library of objects
	including galaxies, planets, lights, static and dynamic objects, items and, others;
	\item Support of player controlled or computer controlled characters.
\end{itemize} 
	


\begin{figure}[p] 
\centering
\begin{subfigure}[t]{.5\textwidth}
    \centering
	\includegraphics[height=2.5in]{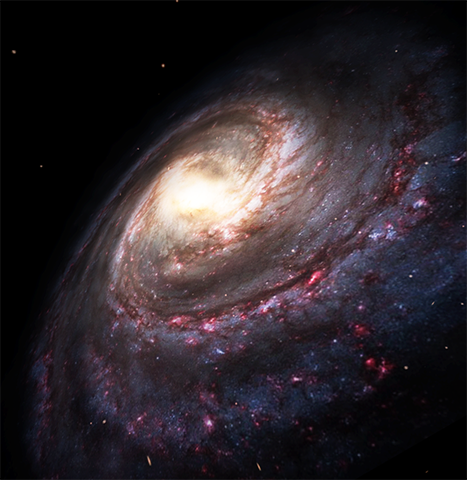}
	\caption{A spiral supergalactic system}
\end{subfigure}%
~ 
\begin{subfigure}[t]{.5\textwidth}
    \centering
	\includegraphics[height=2.5in]{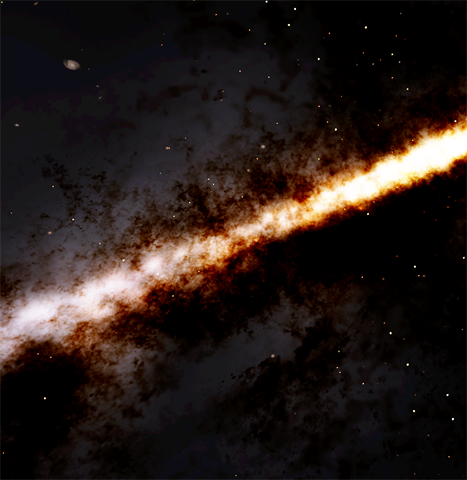}
	\caption{A single galactic}
\end{subfigure} 
 
\begin{subfigure}{.5\textwidth}
    \centering
	\includegraphics[height=2.5in]{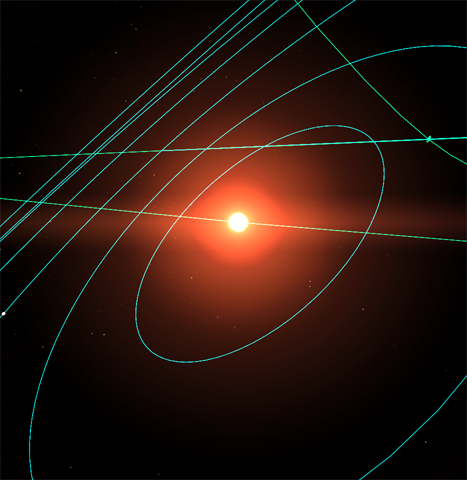}
	\caption{A planetary system}
\end{subfigure}%
~ 
\begin{subfigure}{.5\textwidth}
    \centering
	\includegraphics[height=2.5in]{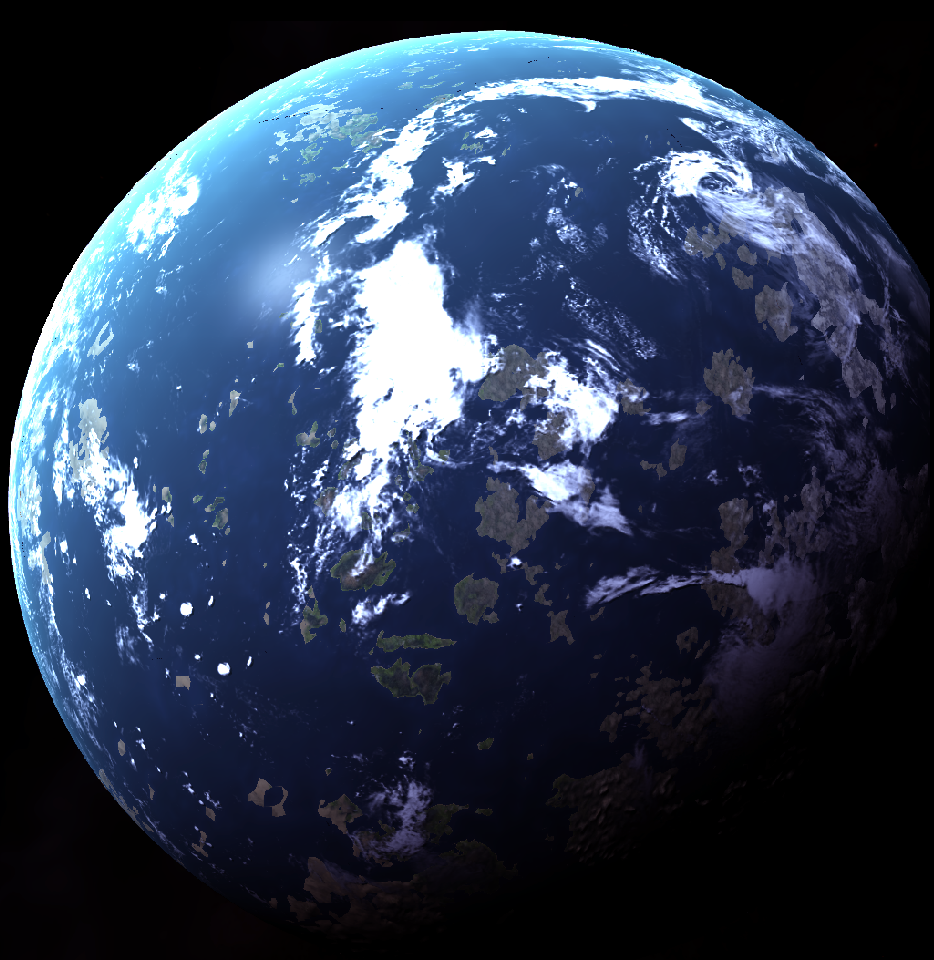}
	\caption{An earth-like planet seen from the orbit}
\end{subfigure}
 
\begin{subfigure}{.5\textwidth}
    \centering
	\includegraphics[height=2.5in]{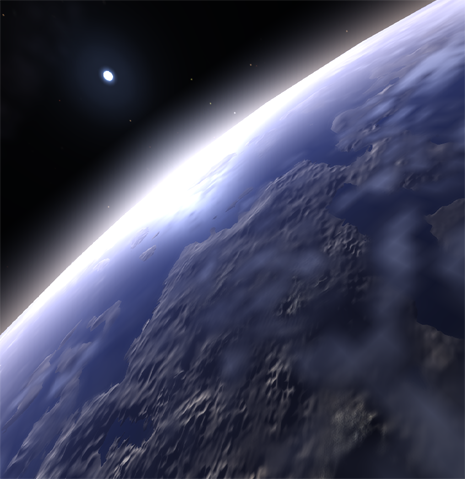}
	\caption{A planet seen from the stratosphere}
\end{subfigure}%
~ 
\begin{subfigure}{.5\textwidth}
    \centering
	\includegraphics[height=2.5in]{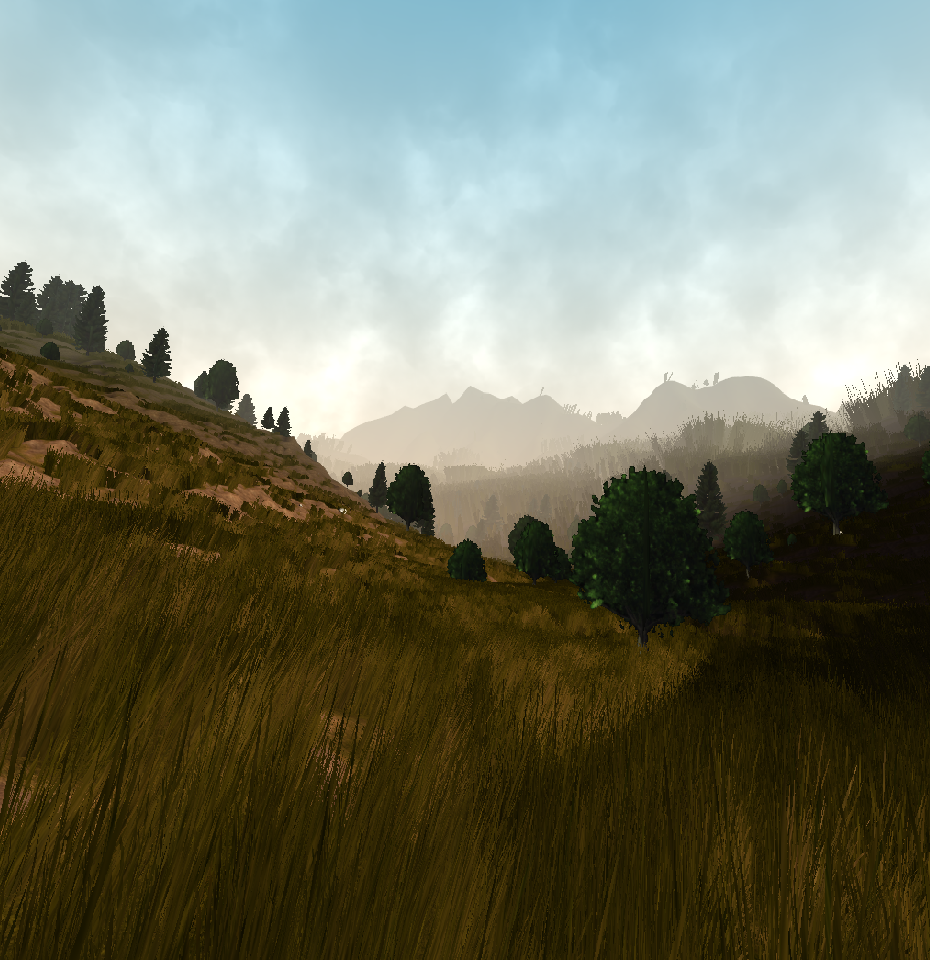}
	\caption{The terrain of a planet}
\end{subfigure}\\ 
  
\caption{The default world hierarchy in SN-Engine environment}
\label{fig:composite}
\end{figure}

The base engine realization contains a procedurally generated universe with
standard node hierarchy~(Fig.~\ref{fig:composite}) ranging from galaxies to planet 
surfaces and their objects. Hierarchy is displayed in columns from left to right and from top to bottom:  
galaxies~(a), spiral galaxy stars and dust~(b), a star system, planets and their orbits~(enabled for clarity)~(c),
an earth-like planet~(d), planet surface seen from a low orbit~(e), planet surface from ground level~(f). 
In addition, other nodal hierarchies can be created
with custom nodes by means of procedural generation algorithms.

A planet has a touchable surface endowed with physical properties which
satisfy the laws of the physics of solids. 
Players can walk on a planet's surface and interact with other objects.

\section{IMPLEMENTATION OF THE GEOMETRIC MODELLING ENGINE} 

The engine is implemented in \CS\, on \textbf{.NET Framework} 
and uses \textbf{Lua}~\cite{Lua} as the main scripting language. The engine 
consists of core systems and modules.

The core systems are as follows:
\begin{itemize}  
	\item Task system managing task creation, processing, linking and balance;
	\item IO system or virtual file system, providing synchronous and 
	asynchronous asset loading, reloading, and dependency management;
	\item Object system or global object management system, 
	base for Assets, Nodes, Components, Events, and others;
	\item Event system that manages subscription and invocation of functions;
	\item Physics system handles physical interactions 
	of nodes with instances of physics entities;
	\item Sound system supports playback of sound effects 
	and ambience in 3D space;
	\item Horizon system providing frame of reference update 
	and loading of nodes;
	\item Input system or user input engine interface;
	\item Net system that enables client-server event relay;
	\item Rendering system which does visualization of nodes 
	with attached cDrawable component;
	\item Scripting system providing safe and real-time 
	game logic programming in \textbf{Lua}.
\end{itemize}  

The list of modules comprises the following:
\begin{itemize}  
	\item Profile which stores user data, settings, saves, etc.;
	\item Add-on or user data package manager;
	\item GUI that renders user interface objects and their rendering;
	\item ModelBuilder providing classes and functions for dynamic model construction;
	\item ProcGen or operation based procedural generation system;
	\item sGenerator or script based procedural generation system;
	\item Flow, the node based visual programming language;  
	\item Forms, the dynamic index database of user defined data assets;
	\item VR, the virtual reality Oculus HMD interface;
	\item Web module that provides Awesomium browser interface, rendering to texture and control functions.
\end{itemize}  

Data modules include primary asset modules which contain methods for asset manipulation in JSON format~\cite{JSON}, 
Material, StaticModel, Particles, Sprite font, SurfaceDataStorage, 
Package, Texture, Sound, and secondary format modules that are used in data import stage: SMD, FBX, BSP, OSM, MCV.

\subsection{Hierarchical Nodal System of the Geometric Modelling Engine}

The core component of the engine world structure is a node. 
Every node follows the set of rules:
\begin{itemize}  
	\item A node can have space bounds with box, sphere or compound shape;
	\item A node can contain any finite number of other nodes and one parent node, forming a tree graph structure;
	\item A node has absolute size variable measured in meters per node unit, so in case of node with spherical 
	bounds its radius is equal to the absolute size of the node;
	\item A node has position, rotation and scale variables which define its location and scale in its parent node; 
	\item A node has seed variable used for procedural generation;
	\item Nodes can have components, custom variables and event listeners;
	\item A node without parent node is the world node.
\end{itemize}

A possible node type hierarchy of a client camera flying 
two meters above the planet surface is as follows:  
0) world\_sol; 1) spacecluster; 2) galaxy; 3) starsystem; 4) star; 5) planet; 6) planet\_surface; 7) planet\_surface\_node; 8) camera.
All types in this hierarchy, with the exception 
for the camera type, are defined in script files.

The location variables and node hierarchy allows 
one to build relative transformation matrices and 
their inverse matrices for each parent node in its 
hierarchy. This in turn allows one to obtain relative 
transformations for any node within the same world.

The precision of 32 bit floating point numbers which are 
widely used in computer graphics is limited while the corresponding 
precision distribution is not linear. Any 
number with 7 or more significant decimal digits 
is subject to data loss during mathematical operations.
There are several workaround methods to overcome this limitation and 
each method needs to calculate object positions in relation 
to camera at runtime.
These methods include the following:
\begin{itemize}
\item Storing object positions as 64bit floats,
	which raises overall precision to 15 decimal digits
	but still has the same precision distribution limitations;
\item Storing object positions as 32bit or 64bit integers,
	since integers have linear precision distribution;
\item Storing object positions in relation to specialized group object.
	This method is used in most geometric modelling systems;
\item Storing relative local object positions with respect to their parent object.
	This method requires hierarchical organization of objects.
\end{itemize}
We use the last method to solve this problem.
This is done in three steps. The first two steps 
are performed when creating or updating a node 
while the third is performed at the rendering stage.

The first step is to compute the node world matrix, which 
consists of the node scale, rotation and position:
\[ \label{weq1}  
	\begin{matrix} \small
	W = \begin{pmatrix}
	S_{x} & 0 & 0 & 0 \\
	0 & S_{y} & 0 & 0 \\
	0 & 0 & S_{z} & 0 \\
	0 & 0 & 0 & 1 \\
	\end{pmatrix}\begin{pmatrix}
	1\  - \ 2(R_{y}^{2} + R_{z}^{2}) & 2(R_{x}R_{y} + R_{z}R_{w}) & 2(R_{z}R_{x} - R_{y}R_{w}) & 0 \\
	2(R_{x}R_{y} - R_{z}R_{w}) & 1\  - \ 2(R_{z}^{2} + R_{x}^{2}) & 2(R_{y}R_{z} + R_{x}R_{w}) & 0 \\
	2(R_{z}R_{x} + R_{y}R_{w}) & 2(R_{y}R_{z} - R_{x}R_{w}) & 1\  - \ 2(R_{y}^{2} + R_{x}^{2}) & 0 \\
	0 & 0 & 0 & 1 \\
	\end{pmatrix}\times\\ \small
	\times\begin{pmatrix}
	1 & 0 & 0 & 0 \\
	0 & 1 & 0 & 0 \\
	0 & 0 & 1 & 0 \\
	P_{x} & P_{y} & P_{z} & 1 \\
	\end{pmatrix}, \\
	\end{matrix}
\]
that is, $W = \mathscale\left( S \right)\cdot\mathrotation\left( R \right)\cdot\mathtranslation(P).$

Here 
$W$ is the node world matrix, 
$S = (S_x,S_y,S_z)$ is the node scaling vector, 
$R = (R_x,R_y,R_z,R_w)$ is the node rotation quaternion, and 
$P = (P_x,P_y,P_z)$ is the node position vector.

The second step is to compute hierarchical matrices using 
world matrices obtained at the previous step and parent nodes:
\[\label{weq2}  
	H_{i} = \left\{ \begin{matrix}
	Id_4, & \mathif\ i = 0, \vspace{0.4em}\\ 
	H_{i - 1}\frac{\text{S}_{i - 1}}{\text{S}_{i}}W_{i}, & \mathif\ i > 0. \\
	\end{matrix} \right.\
\]
Here
\begin{itemize}
\item
  $Id_4$ is the identity matrix of size 4;
\item
  $i$ is the node level in parent hierarchy, where 0 = node, 1 = parent, 
  2 = parent of parent, \ldots{}, $n$ = world node;
\item
  $H_i$ is the $i$-th level node world transformation matrix;
\item
  $S_i$ is the $i$-th level parent node absolute node size in meters;
\item
  $W_i$ is the $i$-th level parent node world matrix.
\end{itemize}

Finally, the third step is the local world matrix computation~(Fig.~\ref{fig:hworld}, where
(a) is the current node, (b) is the target node, (c) is local world matrix and (d) is 
the closest common ancestor for the current and the target nodes). This matrix is computed
as follows:
\begin{figure}[ht!]
\centering{}\includegraphics[width=0.7\textwidth]{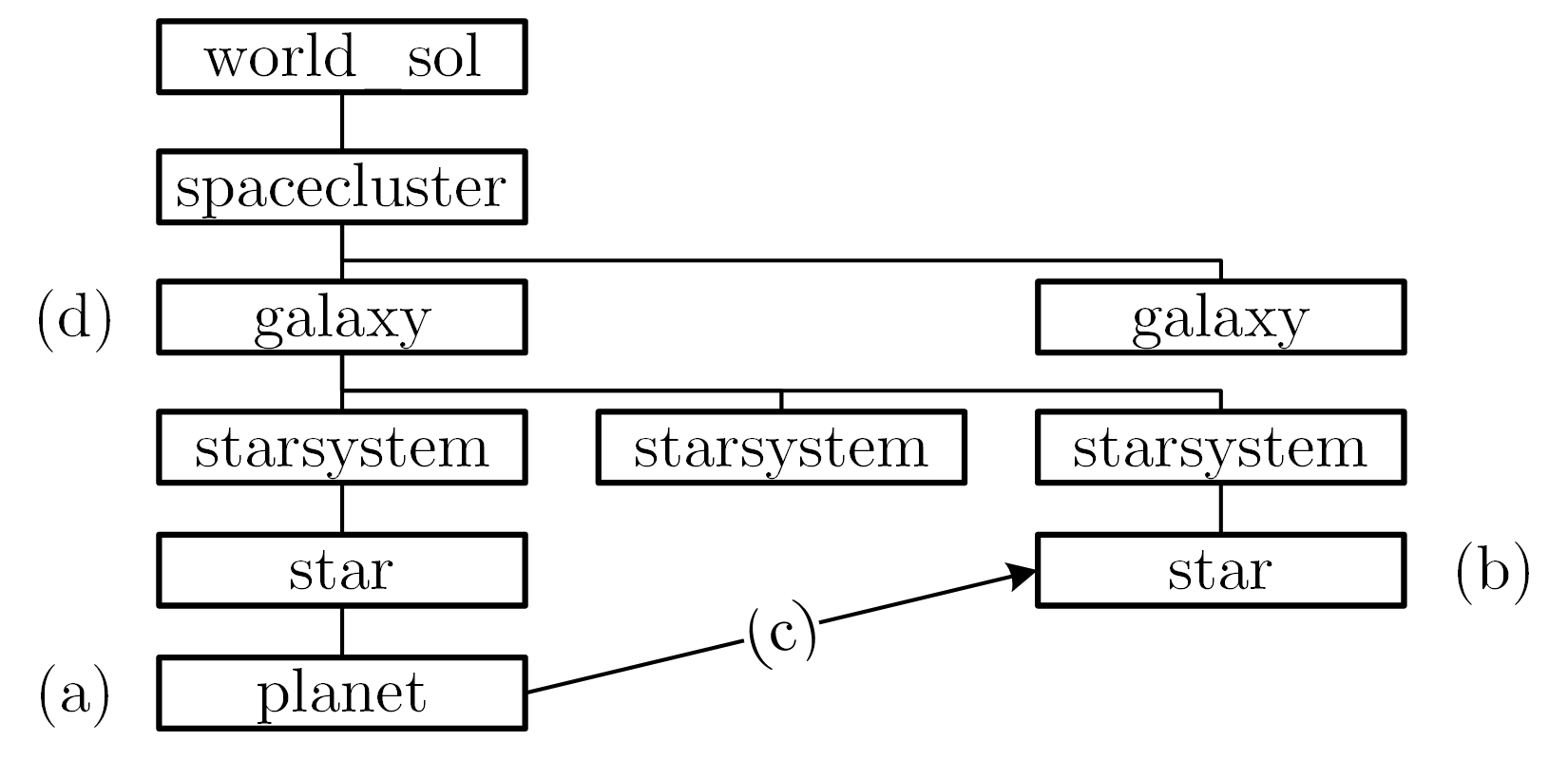}
\caption{Node local world matrix computation} 
\label{fig:hworld}
\end{figure}
\[ \label{weq3}  
	\text{LW}_{N} = H_{\left( C_{\text{level}} - T_{\text{level}} \right)}^{\text{}}H_{\left( N_{\text{level}} - T_{\text{level}} \right)}^{- 1}.
\]
Here 
$LW$ is the local world matrix of node $N$ at node $C$, 
$T$ (the top node) is the nearest node in hierarchy to both current and
target nodes,
$C$ is the current node, 
$N$ is the target node, 
level is the node hierarchy level starting from the world node. 
\vspace{0.5cm}

\subsection{System of the Components}
The engine nodes use partial implementation 
of Entity component system~\cite{ECS} pattern with the 
difference that components can have their own methods. 
The nodes are entities in this pattern. 
Components define how a node interacts 
with other nodes, how it is rendered and how it functions.

Core engine components ({\it sComponent}) include the following: 
Drawable object components ({\it cDrawable}):
{\it cModel} for model instances,
{\it cParticleSystem} for particle system instances,
{\it cSkybox} for skybox instances,
{\it cSurface} for planet surface (terrain and water) instances,
{\it cSpriteText} for text sprites in 3D space,
{\it cAtmosphere} for atmospheric fog,
{\it cVolume} for volumetric texture renderer instances.

Update phase components ({\it cUpdater}): 
{\it cOrbit} which sets node position using orbital parameters and current time,
{\it cConstantRotation} that sets node rotation from current time.

Physics components (include interfaces for physics engine~\cite{BEPU}): 
{\it cStaticCollision} or collision mesh with infinite mass,
{\it cPhysicsSpace} for physical space which contain physics objects,
{\it cPhysicsObject} for physics objects with finite mass and volume, 
{\it cPhysicsMesh} or concave triangulated physics mesh,
{\it cPhysicsCompound} or compound physics object,
{\it cPhysicsActorHull} or physical controller for actors.

Content generation components: 
{\it cRenderer} components that perform render to texture operations,
{\it cCamera} which renders from camera node,
{\it cCubemap} that renders 6-sided cubemap texture,
{\it cHeightmap} draws to planetary height map texture,
{\it cInterface} renders hierarchy of panel objects to GUI texture,
{\it cShadow} draws scene to cascading shadow map~\cite{CSM} texture,
{\it cProcedural} or procedural node generation component.

Other component types:
{\it cLightSource} point light which is used for illumination,
{\it cNavigation} which generates and contains navigational map for AI actors,
{\it cPartition} is hierarchical space partitioning (part of HorisonSystem),
{\it cPartition2D} or quad tree partitioning (4 subspaces on 2 axes),
{\it cPartition3D} or octree partitioning (8 subspaces on 3 axes),
{\it cSurfaceMod} is surface data (height, temperature, etc.) modifier (cSurface),
{\it cWebInterface} is web interface data container.

\subsubsection{Orbital component}
 uses Keplerian orbits to setup its node position. 
When an orbit is created, we calculate the average orbital speed \(V_{\text{orbital}}\)
in terms of of masses and semi-major axis length using the formula 

\[ \label{eq1}
	V_{\text{orbit}\text{al}} = \sqrt{G\frac{M_{\text{node}}M_{\text{parent}}}{a^{3}}}.\
\]
Here 
$G$ is the gravitational constant,
$a$ is the semi-major axis,
$M_{\text{node}}$ is the orbiting node mass,
$M_{\text{parent}}$ is the parent node mass.

Subsequently in the update event we use Algorithm~\ref{alg1} to calculate the vector position 
along the shifted ellipse and then rotates it around 0 with the argument 
of periapsis along the Y axis, inclination along the X axis and ascending 
longitude along the~Y axis.

\begin{algorithm}[H]  
\caption{orbital position update step}\label{alg1} 
\Input{Orbital parameters and current time.}
\Output{Orbital position.}
\BlankLine
\Fn{GetOrbitalPosition($a, b, e, p, i, l, V_{\text{orbital}}, t$)}{ 
	$\widehat{t} = \modulo\left( t V_{\text{orbital}},2\pi \right) - \pi$\;
	${\overline{P}}_{\text{local}} = \left\langle - a\cos{(\text{GetCorrectedT}(e,\widehat{t}))},0,\sign(\widehat{t})\sqrt{\left( 1 - \frac{x^{2}}{a^{2}} \right)b^{2}} \right\rangle$\;
	${\overline{P}}_{\text{parent}} = {\overline{P}}_{\text{local}}\begin{pmatrix}
		\cos p & 0 & \sin p \\
		0 & 1 & 0 \\
		 - \sin p & 0 & \cos p \\
		\end{pmatrix}\begin{pmatrix}
		1 & 0 & 0 \\
		0 & \cos i & - \sin i \\
		0 & \sin i & \cos i \\
		\end{pmatrix}\begin{pmatrix}
		\cos l & 0 & \sin l \\
		0 & 1 & 0 \\
		 - \sin l & 0 & \cos l \\
		\end{pmatrix}$\;
	\KwRet{${\overline{P}}_{\text{parent}}$}
} 
\Fn{GetCorrectedT($e, \widehat{t}$)}{
	$D = 1$\;
	$E = \widehat{t}$\;
	\While{$D > \epsilon$}{ 
		$\widetilde{E} = e\sin E - \widehat{t}$\;
		$D = \abs{\widetilde{E} - E}$\;
		$E = \widetilde{E}$\;
	}
	\KwRet{$E$}
}
\end{algorithm}
Here 
$a$ is the semi-major axis,
$b$ is the semi-minor axis,
$e$ is the eccentricity,
$p$ is the argument of periapsis,
$i$ is the inclination,
$l$ is the ascending longitude,
$t$ is time,
\(\widehat{t}\) is the orbital position. 
The value $\epsilon = 0.001$ in parent node units is used as a precision limiter.

\subsection{The Horizon System}
The horizon system provides dynamic recalculation 
of coordinate systems, space to space node transfer 
and the space partition update. The system checks 
all nodes with enabled space transfer flag against 
the bounds of their parent node and every other node in it. 
When a node is outside of its parent bounds or inside 
of any other node bounds, the system recalculates 
its position, rotation, velocity, and angular velocity 
and then changes its parent node. 

The partition component generates dynamic tree 
structure from partition nodes and calls events 
to its host node. This allows one to create procedural 
generators for three-dimensional (Octree) objects 
like galaxy generation, galactic cluster generation 
and two-dimensional (Quad tree) objects like planet surfaces. 
It~can also be used for any other type of generation with scripting. 

\subsection{Procedural Generation}
Procedural generation is the crucial part of the engine. 
It allows real-time content generation from pre-defined 
patterns and scripts. There are different methods 
of procedural content generation that are implemented 
in the engine including:
\begin{itemize}
	\item Node generation with scripted procedural 
	generators in \textbf{Lua} or \CS\,, e.g. cluster
	and galaxy generation (\CS), star system generation (\textbf{Lua});
	\item Model and node generation with operation 
	based generation system, e.g. models of 
	objects, buildings; generated building interior world nodes and others;
	\item Texture generation from \textbf{Lua} scripts 
	or JSON files;
	\item Texture generation from cCamera and 
	cCubemap node components;
	\item Surface data generation~\cite{TERTOP,TERFRAC,TERGEN,TERFRAC2} from cHeightmap 
	node component, noise functions~\cite{FREQSP} and JSON parameters.
\end{itemize}

\subsubsection{Operation Based Procedural Generation System.}
\begin{figure}[ht!] \centering

\begin{subfigure}{.5\textwidth}
	\centering
	\includegraphics[height=2.4in]{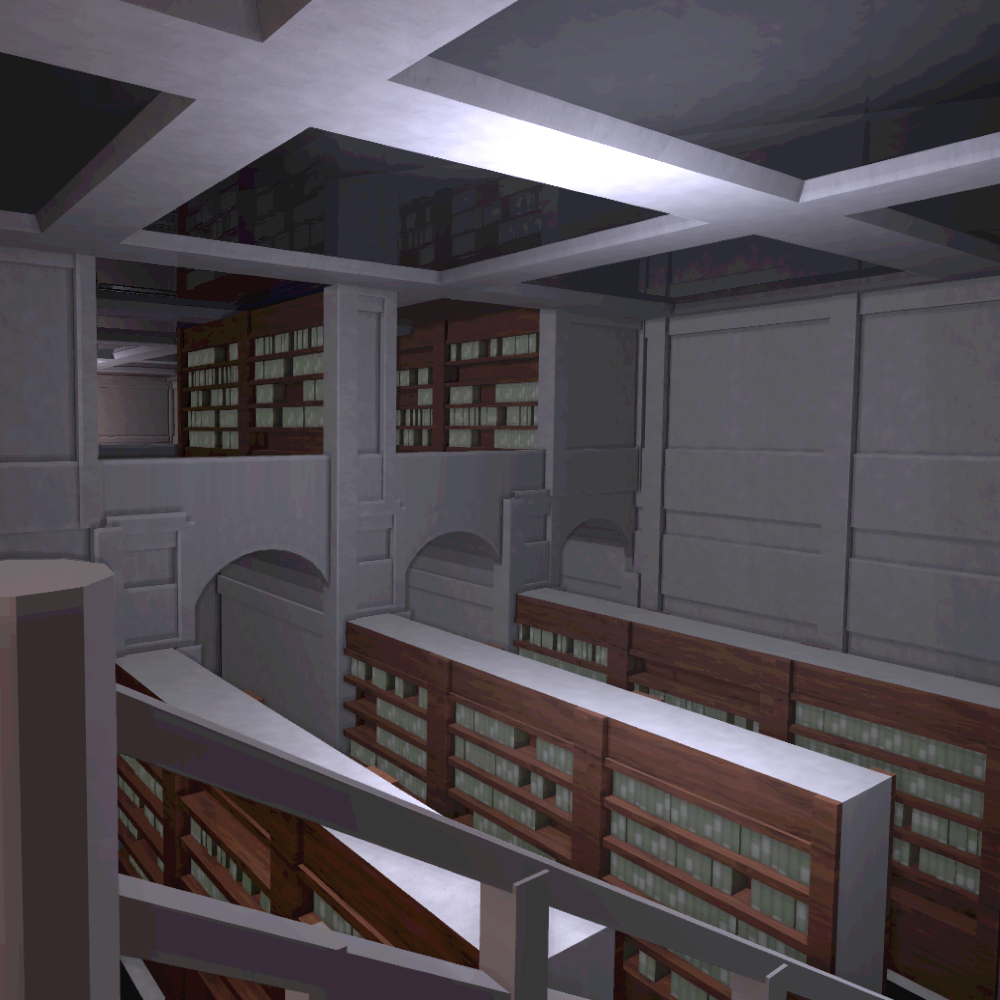}
	\caption{ }
\end{subfigure}%
~
\begin{subfigure}{.5\textwidth}
	\centering
	\includegraphics[height=2.4in]{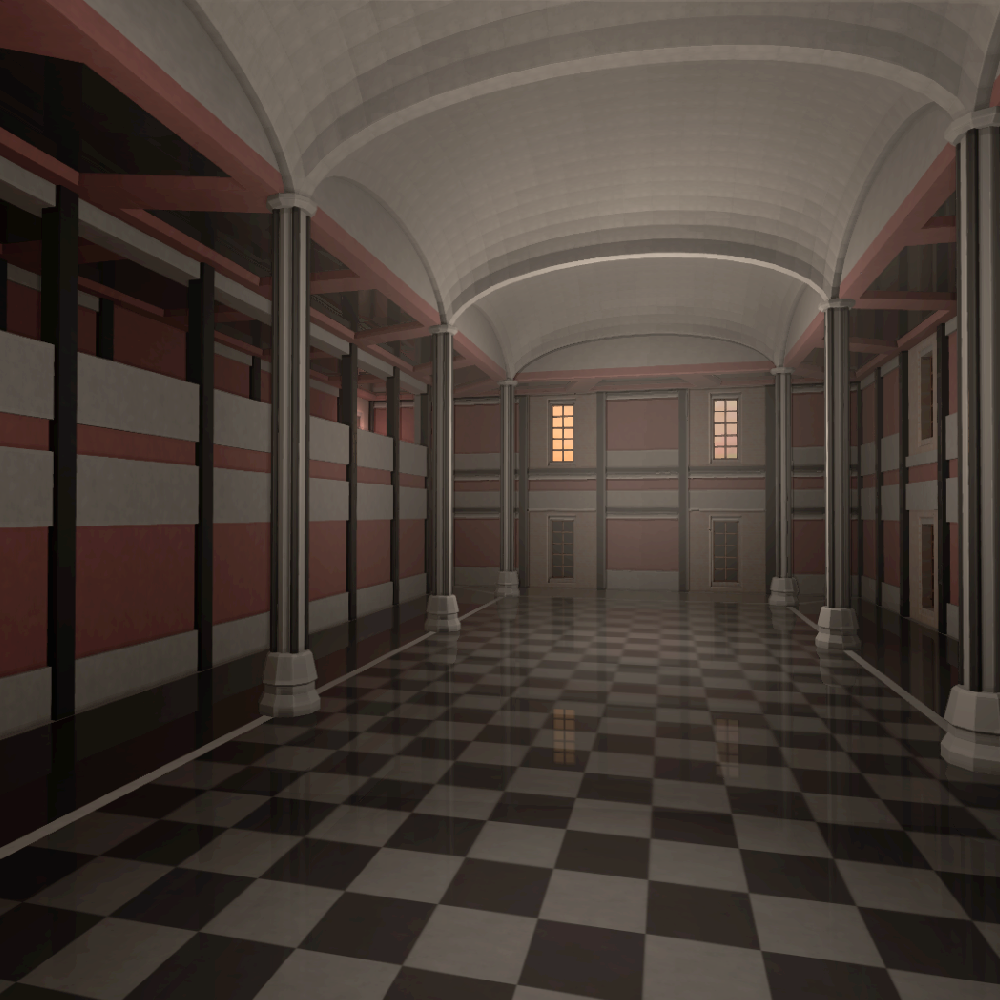}
	\caption{ }
\end{subfigure}

\begin{subfigure}{.5\textwidth}
	\centering
	\includegraphics[height=2.4in]{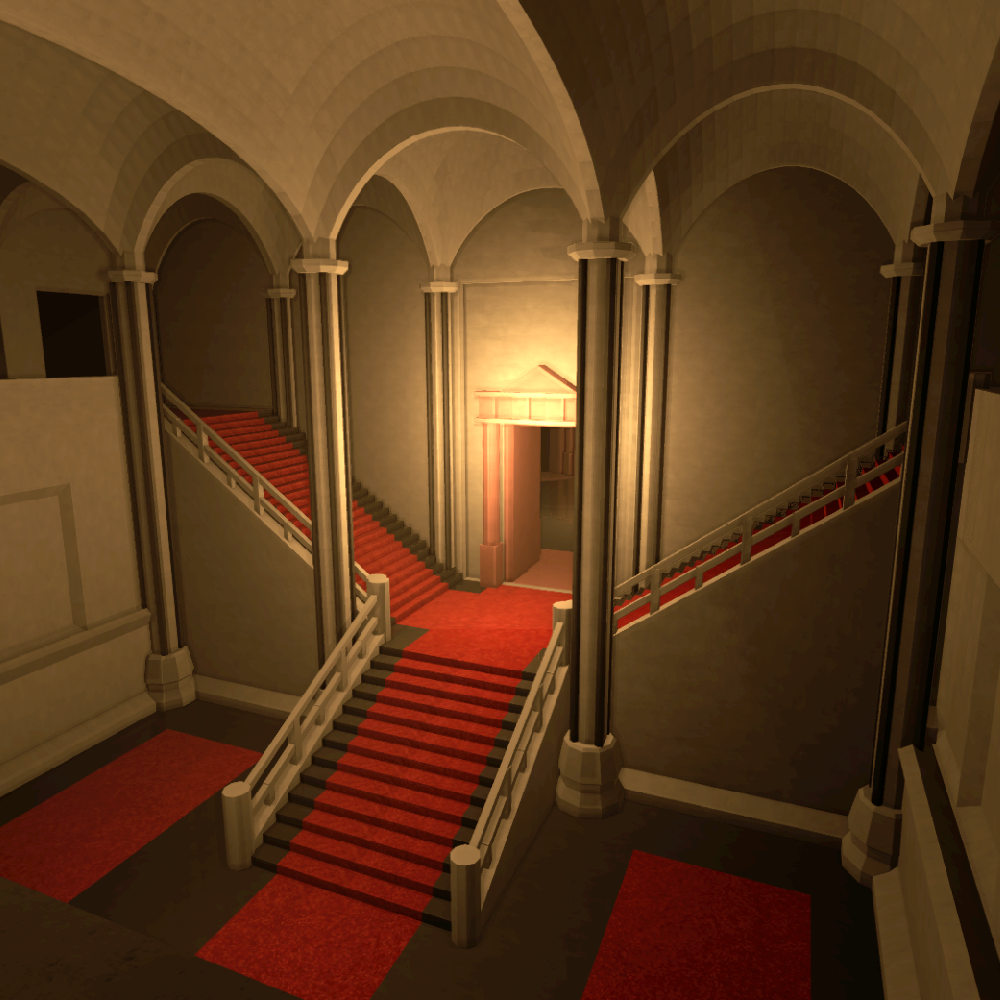}
	\caption{ }
\end{subfigure}%
~
\begin{subfigure}{.5\textwidth}
	\centering
	\includegraphics[height=2.4in]{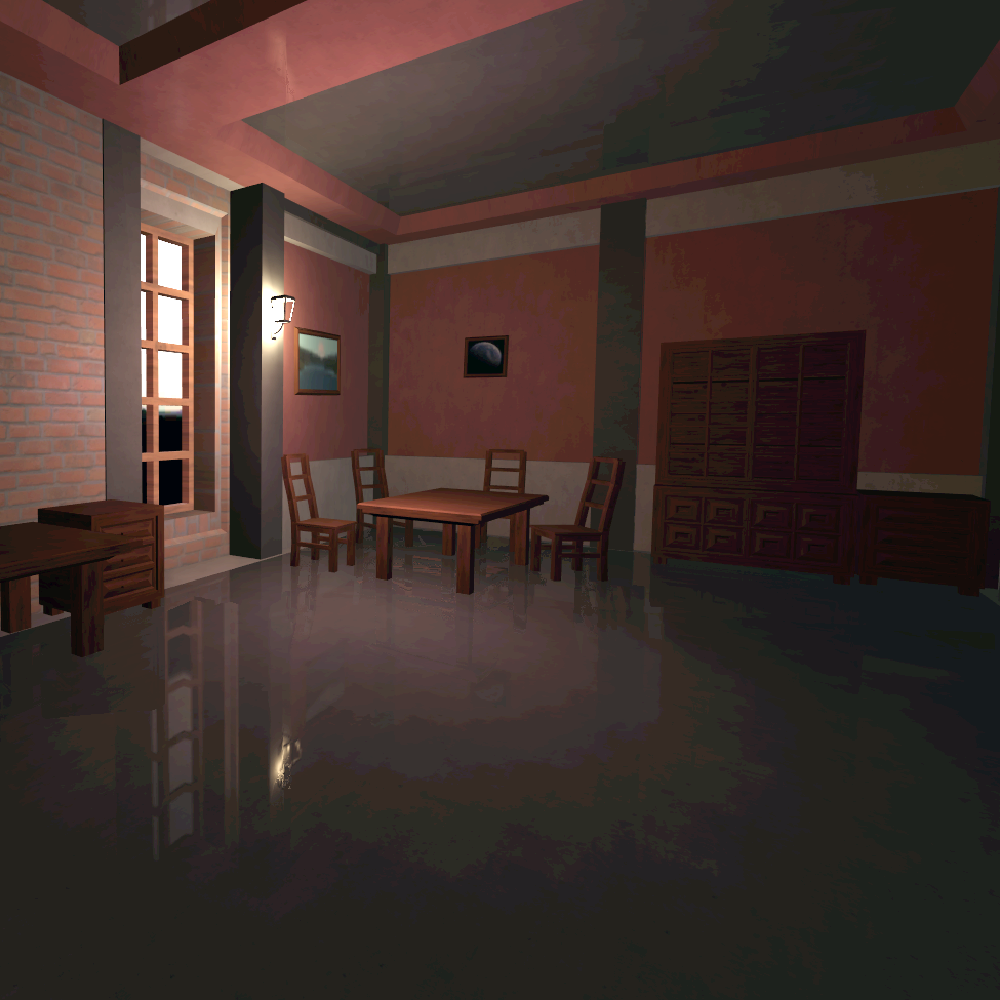}
	\caption{ }
\end{subfigure}
 
\caption{Procedurally generated interiors}
\label{fig:proceduralex}
\end{figure}

This system allows one to define dynamic model structures 
and to build complex 3D models with several levels. 
Models are defined in JSON-like files or 
\textbf{Lua} structures by lists of consecutive operations. 
These operations use named groups of 3D primitives 
along with parameters as input data. Fig.~\ref{fig:proceduralex}
illustrates the use of procedural generation in building
interiors.

The primitive types comprise the following: point as a basic structure which contains position in the local space, 
path as a set of points with a loop flag, 
and surface or polygon which contains edge (path) and other surface properties such as material and uv-matrix.

The procedural generation system includes a total of 56 different 
operation types. The operation types are divided into groups as follows: 
"Create" operations which create sets of primitives from data, 
"Extend" operations which create new primitives from existing, 
"Modify" operations that alter existing primitives, 
"Select" operations which filter existing primitives or return their data, 
and "Utility" operations which provide branching in generation algorithm.

An example of the use of operation is given by
 \{ type : "inset", from:"bt\_base", out : ["bt\_base", "bt\_sides"], extrude : 0.4, amount: 0.5 \}. 
It applies "inset" operation (insets a polygon 
edge by the "amount" value and shifts it by the "extrude" 
value) for each primitive in the group "bt\_base". 
It outputs the central polygon to the "bt\_base" group 
and the edge polygons to the "bt\_sides" group.

\subsection{The Rendering System}
Rendering system uses SharpDX~\cite{SharpDX} library which is an 
open-source managed .NET wrapper of the DirectX API. 
The engine works on DX10 API and DX11 API. 

The system uses a combination of forward and deferred rendering 
techniques~\cite{RENDALL,RENDIL} to draw objects and effects. The rendering 
process is the most difficult one in terms of computational 
complexity and consists of several 
different stages. The preprocessing stage is when all 
texture generation is performed. The rendering system processes 
the incoming draw requests from cRenderer components and fires 
their respective events. Then in the main stage, the 
system performs successive render calls to cDrawable 
components, which are divided by the render groups and 
layers. Finally, at the post-processing stage the system 
combines the results of the previous draw calls into a single 
texture and applies screen effects including 
screen space local reflections~\cite{SSLR} 
and screen space ambient occlusion~\cite{SSAO}.

\subsubsection{Atmosphere rendering shader.}
The atmospheric shader is a~program for GPU which 
manages the rendering of a "fog" layer. This shader
is a work in progress. One of the key algorithms implemented 
in this program is as follows.

\begin{algorithm}[H]  
\caption{Atmosphere glow and fog shader}\label{alg2} 
\Input{A shader parameters $C_{atmosphere}$, $C_i$, $C_{sun}$, $C_{l}$, $M$, $N_i$, 
$N_{planet}$, $H$, $W_{hrz}$, $W_{planet}$, $W_{atmosphere}$, $n$
and back buffer texture $t_{back}$.}
\Output{A pixel colour and alpha values.}
\BlankLine 
\Fn{GetFogColour($N,W_{world}$,$N_{planet},W_{planet},W_{hrz}$,\dots)}{
\tcp*[h]{1. The horizon line and gradient density calculation}

$t_{d}=\mathdot(N,N_{planet})+\frac{W_{hrz}}{W_{planet}}$

$D_{top}=\mathsaturate(t_{d})$

$D_{hrz}=1-\abs{t_{d}}$

$D_{bot}=\mathsaturate(-t_{d})$ 

\tcp*[h]{2. Atmosphere density from view distance calculation}

$D_{world}=1-\mathsaturate(0.01MW_{hrz}W_{world})$

$D_{d}=\mathmax\left(1,\frac{H}{W_{atmosphere}}\right)$ 

\tcp*[h]{3. Light colour and sun glare calculation}

\For{$i=0$ $i<n$ $i++$}{
	$C_{light}\pluseq\mathmax(0, 1-\mathsaturate(5\mathdot(N,N_{i}))C_{atmosphere}C_{i})$
	
	$D_{i}=\mathmax(0,\mathdot(N,N_{i}))$
	
	$S\pluseq(0.5D_{i}+0.4D_{i}^{10}+0.3D_{i}^{100}+0.2D_{i}^{1000})C_{i}$
}

\tcp*[h]{4. Atmosphere parts colour calculation}

$C_{top}= C_{atmosphere} / D_{d}$

$C_{hrz}= C_{haze}\mathlerp(0.25(C_{sun}+C_{light}),C_{light},\mathsaturate(N_{l}+0.1)) / D_{d}+\newline
+ \mathsaturate(N_{h}) / D_{d}$

\tcp*[h]{5. Atmosphere parts density calculation}

$D_{a}=\frac{D_{bottom}+D_{top}}{\mathmax(1,D_{d})}+\frac{D_{hrz}}{D_{d}}$

\tcp*[h]{6. Final atmosphere colour calculation}

\KwRet{$0.5D_{world}D_{a}((D_{hrz}+D_{bot})C_{hrz}+D_{top}(t_{back}+C_{top})+S)$}
}
\end{algorithm}
Here   
$C_{atmosphere}$ is the atmosphere colour, 
$C_i$ is the $i$-th star colour, 
$C_{sun}$ is the star colour, 
$C_{l}$ is the star direction, 
$M$ is the node scale, 
$N$ is the surface normal, 
$N_i$ is the $i$-th star direction, 
$N_{planet}$ is the direction to the planet centre, 
$H$ is the camera distance to the planet surface towards the centre of the planet,
$W_{hrz}$ is the distance to the horizon,
$W_{planet}$ is the distance to the planet centre,
$W_{world}$ is the depth of the current pixel,
$W_{atmosphere}$ is the atmosphere width,
$n$ is the star count,
$t_{back}$ is the pixel colour,
$\mathlerp(a,b,d)$ is the linear interpolation of two vectors $a$ and $b$ based on the weight~$d$.

The resulting image is blended with the current view. The numerical coefficients
in the above formulas have been selected on the basis of visual perception.

\subsubsection{The Scripting System.}
Script system uses \textbf{Lua}~\cite{Lua} as the scripting language. Scripts 
in \textbf{Lua} describe most of game logic. Types of scripts 
include entity definitions, player controller definitions 
(free camera controller, actor controller, etc.), GUI 
widget definitions, auto run scripts and \textbf{Lua} modules 
(definitions for user class types, structure types and 
libraries), procedural generator definitions and others. 
At the time of writing, there were 251 \textbf{Lua} script files in 
the base engine content directory with total size of 904 KB.

\subsubsection{The Networking System.}
The engine network system is based on client-server model 
and uses packets to transfer data. The packets are being sent and 
received asynchronously and incoming packets placed into 
a queue which is then processed from the main thread in 
both server and clients.

\section{LIMITATIONS OF THE PRESENTED GEOMETRIC MODELLING ENVIRONMENT} 
There exist certain limitations for each of the main systems of SN-Engine. 
They are summarized in the next table. 

\begin{table}[ht!] \scriptsize \centering
\caption{Limitations of SN-Engine modelling enviroment} 
\label{tbl:lim} 
\begin{tabular}{|p{2cm}|p{3.6cm}|p{3.4cm}|p{3.4cm}|}
\hline
\textbf{System} & \textbf{Limitations} & \textbf{Approximate value} &
\textbf{Limited by}\\ \hline 
Hierarchical nodal system & maximum node nesting level & \centering 64 & \centering memory
\tabularnewline \cline{2-4}
& maximum loaded node count & \centering tens of thousands & \centering memory
\tabularnewline \cline{2-4}
& maximum absolute node size & $\simeq1\times10^{308}$
& double type precision
\tabularnewline \cline{2-4}
& maximum simultaneous nodes in node & \centering thousands & memory, 
CPU, HDD/SSD read speed
\tabularnewline \hline
Operation based & polygon count & \centering tens of
millions & CPU, int32 maximum value
\tabularnewline \cline{2-4} procedural generation system 
& complex topology & \centering N/A & procedural operations
functionality
\tabularnewline \hline
Rendering system & maximum visible objects & thousands/ millions
instances & GPU memory, GPU processing power
\tabularnewline \cline{2-4}
& maximum drawable objects & \centering tens of thousands & \centering CPU, memory
\tabularnewline \hline
Task system
 & maximum concurrent threads
 & 1 main thread, up to 4 background per CPU core 
 & \centering CPU
\tabularnewline \cline{2-4}
& maximum tasks in thread queue & \centering thousands & \centering CPU, memory
\tabularnewline \cline{2-4}
& maximum concurrent updatable objects & \centering $\simeq500$ &
\centering CPU
\tabularnewline \hline
Physics system
 & maximum active objects in simulation island
 & fast physics: $\simeq100$
slow-motion physics: thousands & \centering CPU
\tabularnewline \cline{2-4}
& maximum static collision triangles per physics space & \centering millions &
memory, CPU
\tabularnewline \hline
Networking system & maximum client connections & \centering tens & CPU, network
bandwidth, memory
\tabularnewline \cline{2-4}
& maximum active networked nodes & \centering hundreds & network
bandwidth
\tabularnewline \hline
IO System & maximum add-on count & \centering hundreds & memory, HDD/SSD read
speed
\tabularnewline \cline{2-4}
& maximum file count & \centering tens of thousands & \centering memory
\tabularnewline \cline{2-4}
 & maximum loaded assets
 & varies from asset filesize
 $\simeq10000-100$
 & \centering memory
\tabularnewline \hline
\textbf{Lua} scripting & processing speed(from \CS) & Divided by 10-1000
& \centering CPU
\tabularnewline \cline{2-4} system
& single threaded. & \centering N/A & \centering CPU
\tabularnewline \hline 
\end{tabular}
\end{table}

\subsubsection{System Requirements.} 
Hardware requirements: 
Processor: Dual Core CPU @ 3.0 GHz, 
Memory: 4GB RAM, 
Hard Drive: 2 GB, 
Graphics: Nvidia GeForce GTX 1070, 2048MB.
Software requirements: 
OS:~Windows 7, 8, 10,
.NET Framework 4.0.


\section{DISCUSSION} 
The new geometric modelling system SN-Engine combines procedural generation algorithms,
arbitrary scale nodal system, and extensive \textbf{Lua} scripting system to construct and 
visualize complex sceneries.

High resolution screenshots and video are available on the system website \newline
at http://snengine.tumblr.com/


\section*{ACKNOWLEDGEMENTS} 
This research was performed in the framework of the state task in the field of scientific 
activity of the Ministry of Science and Higher Education of the Russian Federation, 
project "Development of the methodology and a software platform for the construction 
of digital twins, intellectual analysis and forecast of complex economic systems", 
grant no. FSSW-2020-0008.



\bigskip
\end{document}